\newcommand{\cov}[1]{\mbox{/\hspace{-1.9mm}#1}}
\documentclass[showpacs,preprintnumbers,amsmath,amssymb,nofootinbib]{revtex4}
\usepackage{graphicx}
\usepackage{dcolumn}
\usepackage{bm}
\usepackage{amsfonts}
\usepackage{amssymb}
\usepackage{amsthm}
\usepackage{newlfont}
\begin{document}
\title{\bf Nonlocal Condensate Model for QCD Sum Rules}
\author{Ron-Chou Hsieh$^1$\footnote{E-mail: hsiehrc@phys.sinica.edu.tw}}
\author{Hsiang-nan Li$^{1,2,3,4}$\footnote{E-mail: hnli@phys.sinica.edu.tw}}

\affiliation{$^1$Institute of Physics, Academia Sinica, Taipei
11529, Taiwan}

\affiliation{$^2$Department of Physics, National Tsing-Hua
University, Hsinchu 30013, Taiwan}

\affiliation{$^3$Department of Physics, National Cheng-Kung
University, Tainan 701, Taiwan}

\affiliation{$^4$Institute of Applied Physics, National Cheng-Chi
University, Taipei 11605, Taiwan}

\date{\today}
\begin{abstract}
We include effects of nonlocal quark condensates into QCD sum rules
(QSR) via the K$\ddot{\mathrm{a}}$ll$\acute{\mathrm{e}}$n-Lehmann
representation for a dressed fermion propagator, in which a negative
spectral density function manifests their nonperturbative nature.
Applying our formalism to the pion form factor as an example, QSR
results are in good agreement with data for momentum transfer
squared up to $Q^2 \approx 10 $ GeV$^2$. It is observed that the
nonlocal quark condensate contribution descends like $1/Q^2$,
different from the exponential decrease in $Q^2$ obtained in the
literature, and contrary to the linear rise in the local-condensate
approximation.

\end{abstract}
\maketitle

\begin{figure}[htp]
\centering
\includegraphics[width=0.3\textwidth]{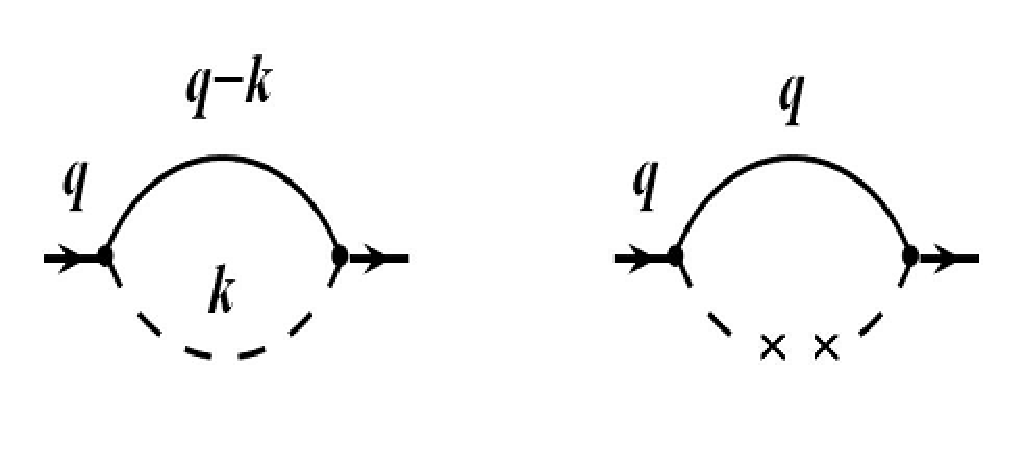}
\caption{Loop diagram with the insertion of the local quark
condensate, where $k$ denotes the loop momentum.}\label{fig1}
\end{figure}

In QCD sum rules (QSR) nonperturbative contributions are taken into
account via vacuum expectation values of nonlocal operators, such as
$\langle \overline{q}(0)q(z)\rangle$ and $\langle G(0)G(z)\rangle$
\cite{SVZ:79a}, where $q$ is a quark field and $G$ is the gluon
field strength. In the standard approach vacuum effects are assumed
to be sufficiently soft to allow the Taylor expansion of, for
instance, the quark condensate $\langle \overline{q}(0)q(z)\rangle$,
at $z=0$ by means of local composite operators,
\begin{eqnarray}\label{eq.1}
\langle \overline{q}(0)q(z)\rangle &=& \langle \overline{q}q\rangle
+ z^{\mu}\langle \overline{q}\partial_{\mu}q\rangle +
\frac{z^{\mu_1}z^{\mu_2}}{2}\langle
\overline{q}\partial_{\mu_1}\partial_{\mu_2} q\rangle + \ldots.
\end{eqnarray}
A local condensate $\langle \overline{q}q\rangle$, i.e., the first
term of the above expansion, prohibits momentum flow. A loop diagram
then turns into a tree diagram as shown in Fig.~\ref{fig1}, when
inserting the local quark condensate into the lower
(nonperturbative) line. The external momentum $q$ flows only through
the upper (perturbative) line, and one has the loop integral
approximated by the product of the propagator $1/q^2$ and the
condensate $\langle \overline{q}q\rangle$ \cite{Ra:0101}. With this
localization assumption, simple hadronic properties including
masses, decay constants, moments of hadronic wave functions, and
form factors have been calculated in QSR.

\begin{figure}[htp]
\centering
\includegraphics[width=0.8\textwidth]{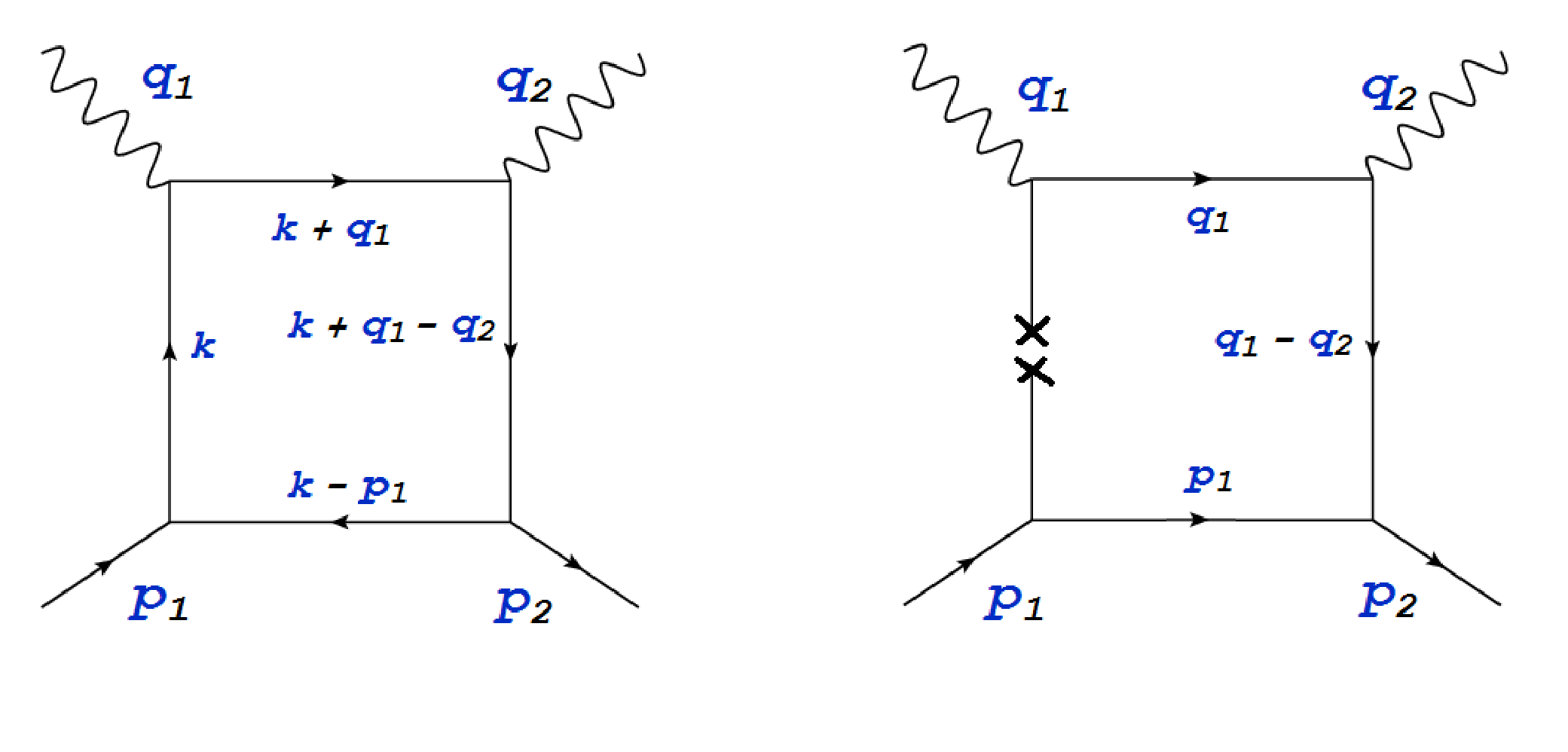}
\caption{Box diagrams for Compton scattering.}\label{fign}
\end{figure}

It has been known that nonperturbative contributions from local
quark condensates grow with the momentum transfer squared $Q^2$ in
form factor calculations, whereas perturbative contributions
decrease \cite{IS:82,NR82}. This is the reason why the standard QSR
approach encounters difficulty, when applied to form factors in the
region with high $Q^2>3$ GeV$^2$ \cite{BPS:09}. It has been observed
that the $Q^2$ dependence of nonperturbative contributions is
moderated by employing the nonlocal quark condensate $\langle
\overline{q}(0)q(z)\rangle$ \cite{BPS:09}. Moreover, using local
quark condensates in QSR analysis of more complicated processes,
such as Compton scattering \cite{CRS93} and the photon structure
function in deeply inelastic scattering \cite{BM94a} which involve
four-point correlation, infrared divergences appear. Consider
the box diagram in Fig.~\ref{fign}, where a light hadron is
scattered by an on-shell photon of momentum $q_1$. The external
momentum $q_1$ flows through the upper horizontal quark line, when
the local quark condensate is inserted into the left vertical quark
line. The upper line then gives a divergent propagator proportional
to $1/q_1^2\to\infty$, and the evaluation of the Wilson coefficient
associated with the quark condensate makes no sense. A resolution of
the above difficulties is to relax the localization assumption.
Including the nonlocal condensates, a finite loop momentum $k$ is
allowed to flow through the box diagram, and the above infrared
divergence is smeared into $1/(q_1+k)^2$. This is our motivation to
investigate effects of the nonlocal quark condensates in QSR. In
this letter we shall set up the framework by studying simpler
processes like the pion form factor, and compare the results with
the local condensates and with the nonlocal condensates.

Nonlocal condensate models \cite{MR:86a} have been applied to QSR
for the pion wave function \cite{BM:95a,MR:89a,MR:92a,Ra:9406},
whose outcome was then treated as an input of the perturbative QCD
factorization formula for the pion form factor \cite{BR:91a}.
Recently, Bakulev, Pimikov and Stefanis calculated the space-like
pion form factor based on QSR with nonlocal condensates
\cite{BPS:09}. They parameterized the nonlocal quark condensate as
$\langle \overline{q}(0)q(z)\rangle=\langle \overline{q}q\rangle
\exp(-|z^2|\lambda_q^2/8)$ \cite{BM98}, where $\lambda^2_q$ is
related to the average virtuality of the condensed quarks. Our
formalism is different, which starts from the
K$\ddot{\mathrm{a}}$ll$\acute{\mathrm{e}}$n-Lehmann (KL)
representation for a dressed propagator of the quark $q$
\cite{IZbook},
\begin{equation}\label{KLform}
\langle\Omega|\mathrm{T}(q(z)\overline{q}(0))|\Omega\rangle =
i\int\frac{d^4k}{(2\pi)^4}e^{-ik\cdot z}\int^\infty_0
d\mu^2\frac{\cov{\it k}\rho_1^q(\mu^2)+\rho_2^q(\mu^2)}{k^2 - \mu^2
+ i\epsilon},
\end{equation}
where $|\Omega\rangle$ represents the exact QCD vacuum, $T$ denotes
the time ordering, the spectral density functions
$\rho_{1,2}^q(\mu^2)$ describe the glutinous medium effect, and
$\mu$ is the effective mass. The KL representation can be deemed as
a superposition of free quark propagators for all mass eigenstates
with the weights $\rho_{1,2}^q(\mu^2)$.

Equation~(\ref{KLform}) is recast into
\begin{equation}\label{arrange}
\langle\Omega|\mathrm{T}(q(z)\overline{q}(0))|\Omega\rangle =
\frac{1}{16\pi^2}\int^\infty_0 ds
\exp\left(\frac{z^2}{4}s\right)\int_0^\infty d\mu^2
\exp\left(-\frac{\mu^2}{s}\right)\left[\frac{i\cov{\it z}
}{2}s\rho_1^q(\mu^2)+\rho_2^q(\mu^2)\right].
\end{equation}
We decompose the above matrix element into the perturbative and
nonperturbative pieces
\begin{eqnarray}
\langle\Omega|\mathrm{T}(q(z)\overline{q}(0))|\Omega\rangle &\equiv&
iZS(z,m_q)+\langle\Omega|:q(z)\overline{q}(0):|\Omega\rangle,
\end{eqnarray}
respectively, with $Z$ being a renormalization constant, $S(z,m_q)$
being the quark propagator in perturbation theory, $m_q$ being the
quark mass. The nonperturbative piece collects the contribution from large $\mu^2$,
\begin{eqnarray}\label{KLformD}
\langle\Omega|:q(z)\overline{q}(0):|\Omega\rangle&=&\frac{1}{16\pi^2}\int^\infty_0
ds\exp\left(\frac{z^2}{4}s\right) \int_{\mu_c^2}^\infty d\mu^2
\exp\left(-\frac{\mu^2}{s}\right)\left[\frac{i\cov{\it z}
}{2}s\rho_1^q(\mu^2)+\rho_2^q(\mu^2)\right].
\end{eqnarray}
The lower bound for the integration variable $\mu^2$ is usually set
to the multi-particle threshold $m_\gamma^2$ in the KL
representation. Here we have modified it into
\begin{equation}
\mu_c^2 =\left\{\begin{array}{cc} c s, & s>m_\gamma^2\\
\\m_\gamma^2, & s\leq m_\gamma^2\end{array}\right.,\label{muc}
\end{equation}
where the free parameter $c$ of order unity will be fixed later.
This modification respects the multi-particle threshold, and at the
same time guarantees a finite integral in Eq.~(\ref{KLformD}). Note
that the integration over $\mu^2$ in Eq.~(\ref{KLformD}) develops a
divergence as the variable $s$ approaches infinity without the above
modification. A negative spectral density function implies
confinement \cite{AS:00a}, and we indeed have the property
$\rho_1^q(\mu^2)<0$ as shown in our formalism below.

We define the distribution functions
\begin{eqnarray}\label{nlc-f}
f_s(s) &=&
\frac{-3}{4\pi^2\langle\overline{q}q\rangle}\int_{\mu_c^2}^\infty
d\mu^2 \exp\left(-\frac{\mu^2}{s}\right)\rho_2^q(\mu^2),\label{fs}\\
f_v(s) &=&
\frac{3}{2\pi^2\langle\overline{q}q\rangle}\int_{\mu_c^2}^\infty
d\mu^2 \exp\left(-\frac{\mu^2}{s}\right)s\rho_1^q(\mu^2),
\end{eqnarray}
and parameterize the spectral density functions as
\begin{eqnarray}\label{spectral}
\rho_1^q(\mu^2) 
=N_1\exp(-a\mu^2)/\mu,\;\;\;\;
\rho_2^q(\mu^2) 
=N_2\exp(-a\mu^2).
\end{eqnarray}
The choice of $\mu_c^2$ in Eq.~(\ref{muc}) then renders the integral in
Eq.~(\ref{fs}),
\begin{eqnarray}
f_s(s)&\propto & \frac{s}{1+a s}\exp(-\mu_c^2/s-a\mu_c^2),
\end{eqnarray}
exhibit the limiting behaviors $\exp(-m_\gamma^2/s)$ at small
$s$ and $\exp(-a c s)$ at large $s$, consistent with
$\exp(-m_\gamma^2/s)$ and the exponential ansatz $\exp(-\sigma_q s)$
postulated in the literature \cite{Ra:9406,BM:0203}, respectively.
Hence, the threshold mass $m_\gamma$ is expected take a value of
order of the constituent quark mass \cite{Ra:91a},
and set to $m_\gamma\sim0.36$ GeV\footnote{There are other choices for the value
of $m_\gamma$, for example, $m_\gamma\simeq0.45$ GeV \cite{Ra:9406},
$m_\gamma\simeq0.50\pm0.07$ GeV\cite{Neu:92a},
$m_\gamma=0.4-0.6$ GeV \cite{BBBD:92a}, etc..} in this work.
Comparing the Taylor expansion of the nonlocal quark condensates
\cite{Gr:95,Ra:0101,BM:0203}
\begin{eqnarray}\label{exactqc}
\langle \overline{q}(0)q(z)\rangle &\equiv&
-\mathrm{Tr}\left[\langle\Omega|:q(z)\overline{q}(0):|\Omega\rangle\right]\nonumber\\
&=& \langle \overline{q}q\rangle \left[1 +
\frac{z^2}{4}\left(\frac{\lambda_q^2}{2} -
\frac{m_q^2}{2}\right)+\cdots\right],\nonumber\\
\langle\overline{q}(0)\gamma_\mu q(z)\rangle &\equiv&
-\mathrm{Tr}\left[\gamma_\mu\langle\Omega|:
q(z)\overline{q}(0):|\Omega\rangle\right]\nonumber\\
&=&-i\frac{z_\mu}{4}\langle \overline{q}q\rangle\left(m_q
+\cdots\right).
\end{eqnarray}
with Eq.~(\ref{KLformD}), we have the constraints
\begin{eqnarray}\label{constrain}
& &\int^{\infty}_0 f_s(s) ds = 1,\;\;\;\;
\int^{\infty}_0 s f_s(s) ds =
\frac{1}{2}(\lambda_q^2 - m_q^2),\nonumber\\
& &\int^{\infty}_0 f_v(s) ds = m_q,
\end{eqnarray}
which determine the free parameters $a$, $N_1$ and $N_2$ in
Eq.~(\ref{spectral}), given values of $\lambda_q$ and $m_q$.

\begin{figure}[htp]
\centering
\includegraphics[width=1\textwidth]{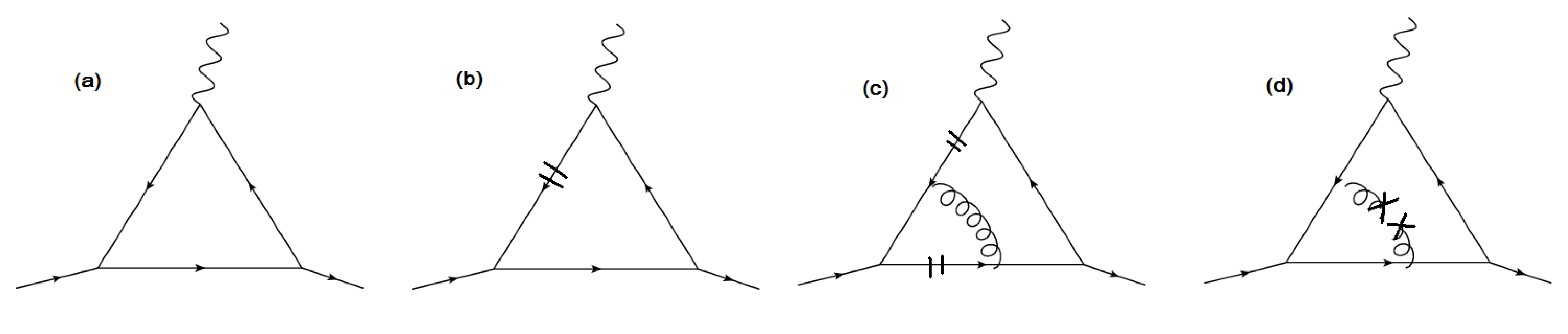}
\caption{(a) Perturbative contribution, (b) two-quark condensate
contribution, (c) four-quark condensate contribution, and (d) gluon
condensate contribution to the pion form factor.}\label{pion}
\end{figure}

The dressed propagator includes both the perturbative and
nonperturbative contributions,
\begin{eqnarray}\label{qpropagator}
S^q(p) &=& \frac{\cov{\it p}+m_q}{p^2-m_q^2} -
\frac{1}{2}i\frac{(\gamma^\alpha\cov{\it p}\gamma^\beta
G_{\alpha\beta}-m_q\gamma_\alpha
G^{\alpha\beta}\gamma_\beta)}{(p^2-m_q^2)^2}\nonumber\\ &-&
\frac{\pi \alpha_s\langle G_{\alpha\beta}^2\rangle m_q\cov{\it
p}(m_q+\cov{\it p})}{(p^2-m_q^2)^4} + \left[\cov{\it p}\hat{I}_1^q +
\hat{I}_2^q\right]\frac{\exp[c(p^2 - \mu^2)/\mu^2]}{p^2-\mu^2},
\end{eqnarray}
with the definitions
\begin{eqnarray}
\hat{I}_{1,2}^q f(\mu) &\equiv& \int_{\mu_c^2}^\infty d\mu^2
\rho_{1,2}^q(\mu^2)f(\mu).
\end{eqnarray}
The second and third terms on the right hand side of
Eq.~(\ref{qpropagator}) arise from the background gluon field
\cite{Ma:84a,Co:93a}, and the forth term comes from the nonlocal
quark condensates with the integrations over $\mu^2$ and $s$ being
exchanged in Eq.~(\ref{KLformD}). As stated before, local quark
condensates lead to contributions linear in $Q^2$, which are more
serious than the constant contributions from local gluon condensates
at large $Q^2$ \cite{IS:82,NR82}. Gluon condensate contributions to
the pion form factor are actually negligible. The contribution
from the quark-gluon-antiquark condensate $\bar{q}Gq$ is smaller than
that from Fig.~\ref{pion}(b) in our model, which is less than 5\% of the
four-quark condensate contribution. Therefore, only the nonlocal quark
condensates are taken into account here.

Inserting Eq.~(\ref{qpropagator}) into the triangle diagrams for the
three-point correlation function, we derive the perturbative and
nonperturbative contributions to the pion form factor $F_\pi(Q^2)$,
\begin{equation}\label{mypionff}
-f_{\pi}^2 F_{\pi}(Q^2)\exp\left(-\frac{2m_\pi^2}{M^2}\right) =
\frac{1}{\pi^2}\left\{\int_0^{s_0}ds_1 ds_2
\rho^{pert}(s_1,s_2,Q^2)\exp\left(-\frac{s_1 +
s_2}{M^2}\right)+\Delta^{quark}+\Delta^{gluon}\right\}.
\end{equation}
In the above expression $f_\pi$ is the pion decay constant, $m_\pi$
is the pion mass, $M$ is the Borel mass, and $s_0$ is the duality
interval. The calculation of the spectral function $\rho^{pert}$
associated with the perturbative contribution, and of the quark
(gluon) condensate contribution $\Delta^{quark}$ ($\Delta^{gluon}$)
involves four types of diagrams displayed in Fig.~\ref{pion}. The
perturbative spectral function and the gluon condensate contribution
are given by \cite{IS:82,NR82}
\begin{eqnarray}
\rho^{pert} &=& \frac{N_c(e_d-e_u)}{2\lambda^{7/2}}Q^4\left\{
s_1(Q^2+s_1)^3+s_2(Q^2+s_2)^3 - s_1s_2[2Q^4+
Q^2(s_1+s_2)-2(s_1^2+s_2^2)+6s_1s_2]\right\},\\
\Delta^{gluon}&=& -\frac{\alpha_s}{12\pi M^2}\langle
G_{\alpha\beta}^2\rangle,\label{gluon}
\end{eqnarray}
respectively, with $N_c$ being the number of colors, $e_u$ ($e_d$) being
the charge of the $u$ ($d$) quark, and the variable
\begin{eqnarray}
\lambda&=&(s_1+s_2+Q^2)^2-4s_1 s_2.
\end{eqnarray}

We compute the quark condensate contribution, obtaining
\begin{equation}\label{cond}
\Delta^{quark} = \langle\overline{q}q\rangle \int_0^{s_0}ds_1 ds_2
\left[(e_u\hat{I}_1^u - e_d\hat{I}_1^d)\rho^{a}_{2qc} +
(e_u\hat{I}_1^d -
e_d\hat{I}_1^u)\rho^{b}_{2qc}\right]\exp\left(-\frac{s_1 +
s_2}{M^2}\right) + \alpha_s\langle
\overline{q}q\rangle^2\Delta^{quark}_{4qc}.
\end{equation}
The two-quark condensate spectral functions $\rho^{a,b}_{2qc}$ and
the four-quark condensate function $\Delta^{quark}_{4qc}$ are
written as
\begin{eqnarray}
\rho^{a}_{2qc} &=& \frac{N_c}{\lambda^{7/2}}
\left[Q^6\mu^2(Q^2+\mu^2)^2 -(Q^4-Q^2\mu^2+\mu^4)(s_1^4+s_2^4)
- Q^2(3Q^4+2Q^2\mu^2-5\mu^4)(s_1^3+s_2^3)\right.\nonumber\\
& & - 3Q^2(Q^6+2Q^4\mu^2-2\mu^6)(s_1^2+s_2^2)-
Q^4(Q^6+2Q^4\mu^2+4Q^2\mu^4+3\mu^6)(s_1+s_2) \nonumber\\
& & - 2(Q^4+2Q^2\mu^2-2\mu^4)(s_1^3 s_2+s_2^3 s_1) +
Q^2(Q^4+2Q^2\mu^2-5\mu^4)(s_1^2 s_2+s_2^2 s_1)\nonumber\\
& & + \left. 6(Q^4+Q^2\mu^2-\mu^4)s_1^2 s_2^2 -
2Q^2(Q^6+5Q^4\mu^2-2Q^2\mu^4-6\mu^6)s_1 s_2\right],\nonumber\\
\rho^{b}_{2qc} &=& -\frac{N_c Q^4}{2\lambda^{7/2}}
\left[2Q^2\mu^2(3Q^4-12Q^2\mu^2+10\mu^4) +
(s_1^4+s_2^4) + 3(Q^2-2\mu^2)(s_1^3+s_2^3) \right.\nonumber\\
& & + 3(Q^4-2Q^2\mu^2+2\mu^4)(s_1^2+s_2^2) +
Q^2(Q^4+6Q^2\mu^2-18\mu^4)(s_1+s_2)\nonumber\\
& & \left. + 2(s_1^3 s_2+s_2^3 s_1) - (Q^2-6\mu^2)(s_1^2 s_2+s_2^2
s_1) - 6s_1^2 s_2^2 - 2(Q^4-12Q^2\mu^2+6\mu^4)s_1 s_2\right],\\
\Delta^{quark}_{4qc} &=& (e_u\hat{I}_2^u - e_d\hat{I}_2^d)
\lim_{m^2\rightarrow0}\frac{\partial}{\partial m^2}
\left(\int^{s_0}_{m^2}ds_1 \int^{s_0}_{\alpha}ds_2 +
\int_0^{m^2}ds_1 \int_0^{\alpha}ds_2\right)
(e^{-s_1/M^2}-1)g_1e^{-s_2/M^2} \nonumber\\ &+& (e_u\hat{I}_2^d -
e_d\hat{I}_2^u) \lim_{m^2\rightarrow0}\frac{\partial}{\partial m^2}
\left(\int^{s_0}_{\mu^2}ds_1 \int^{s_0}_{\beta}ds_2 +
\int_0^{\mu^2}ds_1
\int_0^{\beta}ds_2\right)(e^{-s_2/M^2}-1)g_2e^{-s_1/M^2},\label{19}
\end{eqnarray}
with the functions
\begin{eqnarray}
g_1 &=& \frac{8\pi
N_c}{3\lambda^{5/2}s_1}\left\{Q^4[-6m^4-\mu^2(Q^2+\mu^2)+m^2(4Q^2+6\mu^2)]
+ (Q^2+\mu^2)s_1^3+(Q^2-\mu^2)s_2^3 \right.\nonumber\\
& & - (2m^2Q^2-2Q^4-3Q^2\mu^2+\mu^4)s_2^2 -
(2m^2Q^2-2Q^4-Q^2\mu^2+\mu^4)s_1^2\nonumber\\ & & +
Q^2[Q^4-Q^2\mu^2-2\mu^4+2m^2(Q^2+3\mu^2)]s_1 +
Q^2[Q^4+3Q^2\mu^2+4\mu^4+2m^2(Q^2-3\mu^2)]s_2\nonumber\\ & & \left.-
(Q^2+3\mu^2)s_1^2s_2 - (Q^2-3\mu^2)s_2^2s_1 +
2s_1s_2(2m^2Q^2-Q^4-2Q^2\mu^2+\mu^4)\right\},\nonumber\\
g_2 &=&\frac{8\pi
N_c}{3\lambda^{5/2}s_2}\left\{Q^4[Q^4+m^4-6\mu^2(Q^2-\mu^2)+2m^2(Q^2-3\mu^2)]
+ (m^4-4m^2Q^2+Q^4)s_1^2+(Q^2+m^2)^2s_2^2 \right.\nonumber\\ & & +
2Q^2s_2(Q^2+m^2)(Q^2+m^2-3\mu^2) -
2Q^2s_1[2m^4-(Q^2-m^2)(Q^2-3\mu^2)]\nonumber\\ & & \left.
-2s_1s_2(m^4-m^2Q^2-2Q^4)\right\},
\end{eqnarray}
and the variables
\begin{eqnarray}
\alpha &=&
(m^2Q^2+\mu^2s_2)\left(\frac{1}{Q^2+\mu^2}+\frac{1}{s_2-m^2}\right),
\nonumber\\ \beta &=&
(\mu^2Q^2+m^2s_1)\left(\frac{1}{Q^2+m^2}+\frac{1}{s_1-\mu^2}\right).
\end{eqnarray}
Note that the singularity from $s_1\to 0$ ($s_2\to 0$) in the
function $g_1$ ($g_2$) is removed by the factor $(e^{-s_1/M^2}-1)$
[$(e^{-s_2/M^2}-1)$] in Eq.~(\ref{19}). It is observed that the
contributions from the nonlocal quark condensates must be power-like
in $Q^2$ in the asymptotic limit, no matter how to parameterize
$\rho_{1,2}^q(\mu^2)$. The dominant contribution
$\Delta^{quark}_{4qc}$ descends like $1/Q^2$ as $Q^2\to\infty$,
which is different from the exponential decrease in $Q^2$ obtained
in \cite{BPS:09}, and contrary to the linear rise in the local
condensate approximation \cite{IS:82,NR82}.

A remark is in order. As calculating condensate contributions in the
conventional QSR approach, the upper bound of the integration variable
$s$ is usually extended to infinity. In our formalism both the perturbative and
condensate contributions are calculated in the same framework with the
dressed quark propagators. Hence, it is more natural to parameterize the
continuum contribution to the spectral function on the hadronic side of
the sum rule as that on the operator-product-expansion side for $s>s_0$,
which includes the condensate terms. After cancelling the continuum
contributions from both sides of the sum rule, the upper bound $s_0$
appears in Eq.~(\ref{cond}). This is a difference between our formalism and
the conventional QSR approach.

\begin{figure}[htp]
\centering
\includegraphics[width=0.4\textwidth]{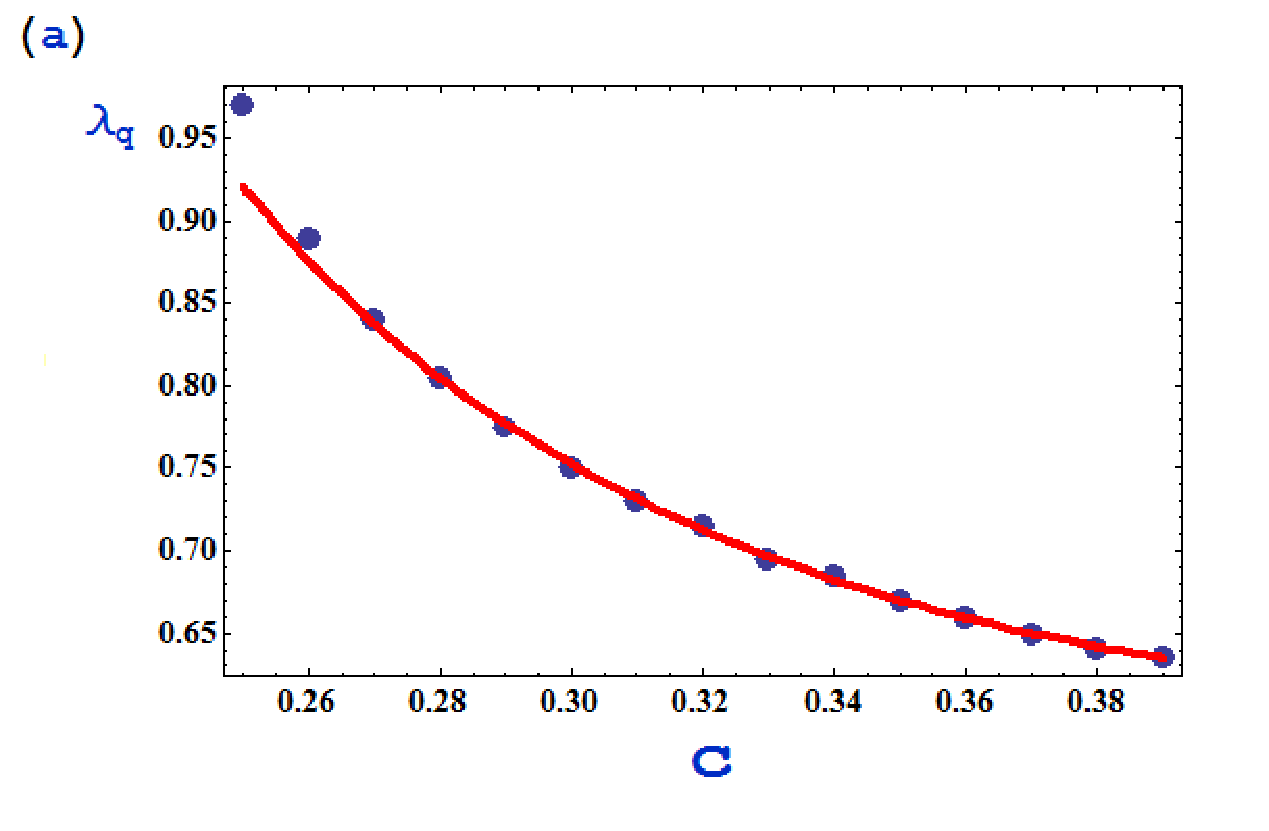}

\includegraphics[width=0.4\textwidth]{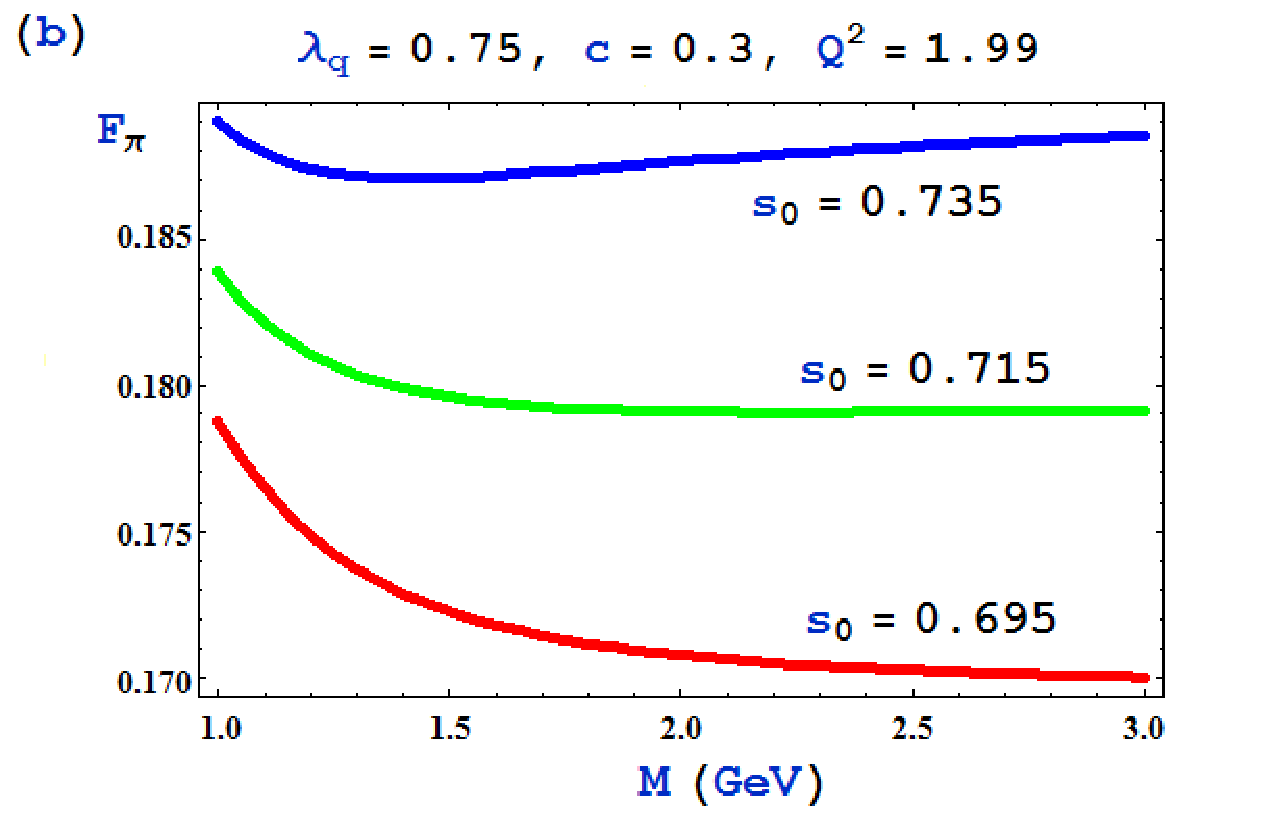}
\includegraphics[width=0.4\textwidth]{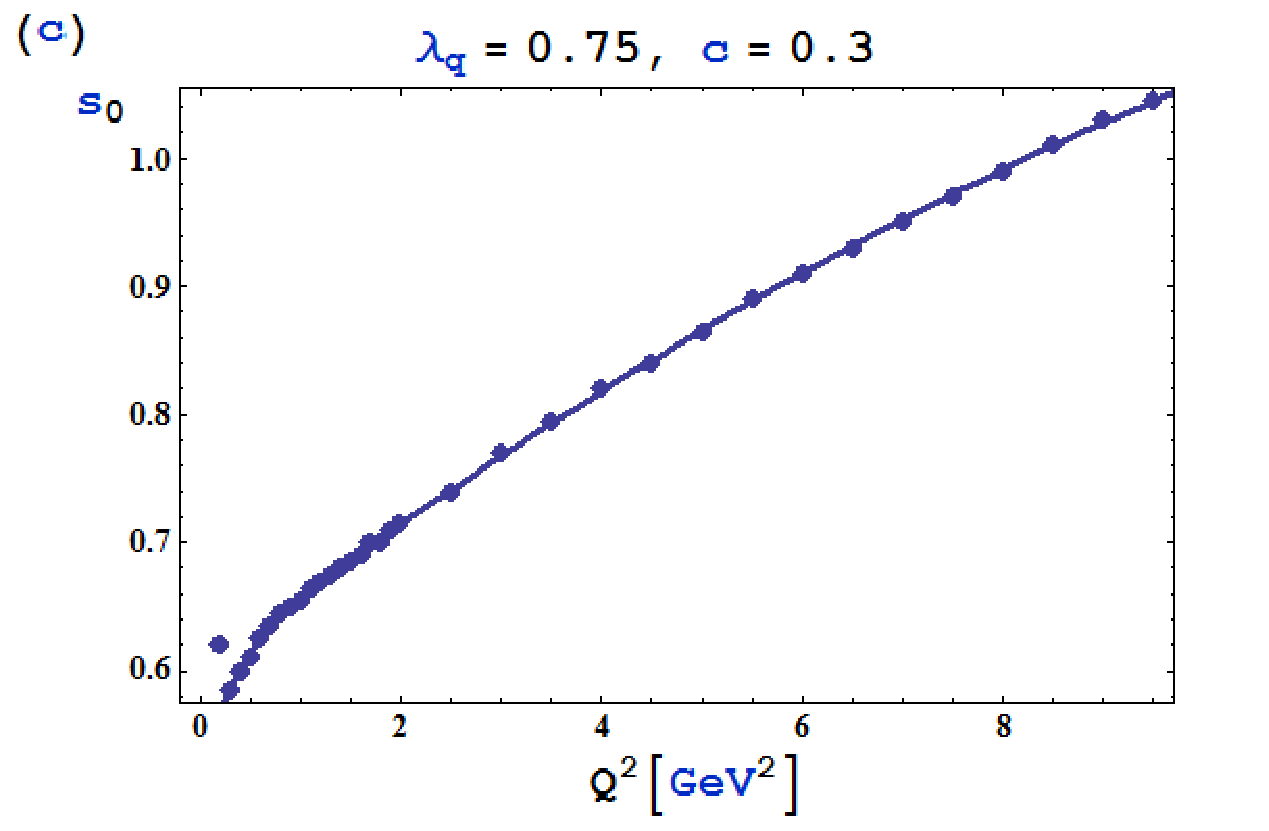}
\caption{(a) Curve for the allowed values of $c$ and $\lambda_q$ that
produce the data of $F_\pi(Q^2=1.99\;\mathrm{GeV}^2)$. (b)
$M$ dependence of $F_\pi(Q^2=1.99\;\mathrm{GeV}^2)$ for different
$s_0$ with $\lambda_q=0.75$ GeV and $c=0.3$. (c) $Q^2$ dependence of
$s_0$ for the pion form factor. The function of the fitting curve is
presented in Eq.~(\ref{curve}).}\label{s0choice}
\end{figure}

The local condensates appearing in Eqs.~(\ref{gluon}) and
(\ref{cond}) are taken to be \cite{Io:05}
\begin{eqnarray}
\frac{\alpha_s}{\pi}\langle G_{\alpha\beta}^2\rangle &=&
0.005\pm0.004 \;\;\mathrm{GeV}^2, \nonumber\\ \langle
\overline{q}q\rangle &=&
-(1.65\pm0.15)\times 10^{-2}\;\; \mathrm{GeV}^3, \nonumber\\
\alpha_s\langle \overline{q}q\rangle^2 &=& (1.5\pm0.2)\times 10^{-4}
\;\;\mathrm{GeV}^6.
\end{eqnarray}
The duality interval $s_0(Q^2)$ at a given $Q^2$ is determined by
the requirement that the form factor is least sensitive to the Borel
mass $M$. The average virtuality $\lambda_q$ and lower bound c,
being not known with certainty, are fixed by fits to the data of the
pion form factor $F_\pi(Q^2)= 0.179\pm0.021$ at $Q^2=1.99$ GeV$^2$
\cite{BBHk:78a,Vol:01a,Hor:06a,Tad:06a}. In figure~\ref{s0choice}(a)
we display the allowed values of $c$ and $\lambda_q$ as a curve in
the $c$-$\lambda_q$ plane. The range of $\lambda_q$ is consistent
with $\lambda_q=0.63$ GeV from QSR \cite{BI82} and $\lambda_q=0.85$
GeV from the instanton analysis \cite{PW:96}. Below we adopt
$\lambda_q=0.75$ GeV and $c=0.3$ to produce the central values of
our predictions for the pion form factor. Choosing the light quark
masses $m_u=4.2$ MeV and $m_d=7.5$ MeV, we solve for the free
parameters $a$, $N_1$ and $N_2$ from the constraints in
Eq.~(\ref{constrain}), whose results are listed in
Table~\ref{unknown}. The product $ac\approx 6.8$ GeV$^{-2}$ is in
agreement with the value of $\sigma_q\approx 10$ GeV$^{-2}$
postulated in \cite{DEM97}. The opposite signs of $N_1$ and $N_2$
imply the violation of positivity, which can be interpreted as a
manifestation of confinement \cite{AS:00a}.

\begin{table}[htp]
\centering
\begin{tabular}{|@{\hspace{0.3cm}}c@{\hspace{0.3cm}}|@{\hspace{0.3cm}}c@{\hspace{0.3cm}}|
@{\hspace{0.3cm}}c@{\hspace{0.3cm}}|@{\hspace{0.5cm}}c@{\hspace{0.5cm}}
|@{\hspace{0.5cm}}c@{\hspace{0.5cm}}|@{\hspace{0.5cm}}c@{\hspace{0.5cm}}
|} \hline
 &  $\lambda_q$ (GeV) & $m_q$ (MeV) & $a$ (GeV$^{-2}$)
 & $N_1/\langle \overline{q}q\rangle$ (GeV$^{-4}$)& $N_2/\langle \overline{q}q\rangle$ (GeV$^{-4}$)\\
\hline
$u$ quark &  $0.75$ & $4.2$ & $22.7$ & $20.54$ & $-7784.54$ \\
\hline
$d$ quark &  $0.75$ & $7.5$ & $22.7$ & $36.70$ & $-7789.33$   \\
\hline
\end{tabular}
\caption{Parameters associated with the quarks $u$ and $d$ in our
formalism. } \label{unknown}
\end{table}


Figure~\ref{s0choice}(b) indicates the best choice of $s_0=0.715$
GeV$^2$ with $\lambda_q=0.75$ GeV and $c=0.3$, at which the pion
form factor $F_\pi(Q^2=1.99\;\mathrm{GeV}^2)$ becomes independent of
$M$ for $M>1.5$ GeV. In the calculation below, we simply set the
Borel mass to $M=1.5$ GeV. In Fig.~\ref{s0choice}(c) we present the
$Q^2$ dependence of the best choice $s_0(Q^2)$ with the same inputs,
whose curve is close to a straight line:
\begin{eqnarray}
s_0(Q^2)=0.6+0.06Q^2-0.0014Q^4, \label{curve}
\end{eqnarray}
for $Q^2>1$ GeV$^2$. It is seen that $s_0$ drops rapidly in the
region of low $Q^2<1$ GeV$^2$, where QSR are supposed to be
inapplicable. $s_0$ in Fig.~\ref{s0choice}(c), increasing from 0.65
GeV$^2$ to 1.05 GeV$^2$ for 1 GeV$^2<Q^2<$ 10 GeV$^2$, shows a bit
stronger $Q^2$ dependence compared to that in \cite{BPS:09}.
Nevertheless, its range obeys the postulation \cite{BPS:09} that it
should not be lower than the middle point 0.6 GeV$^2$ of the
interval between the meson masses $m_\pi^2 = 0$ and $m^2_{A_1}= 1.6$
GeV$^2$. Besides, we have confirmed that the pion decay constant
squared takes the value $f_\pi^2\simeq0.0171$ GeV$^2$ for
$s_0\approx 0.7$ GeV$^2$ in our formalism with the nonlocal quark
condensates.

\begin{figure}[htp]
\centering
\includegraphics[width=0.4\textwidth]{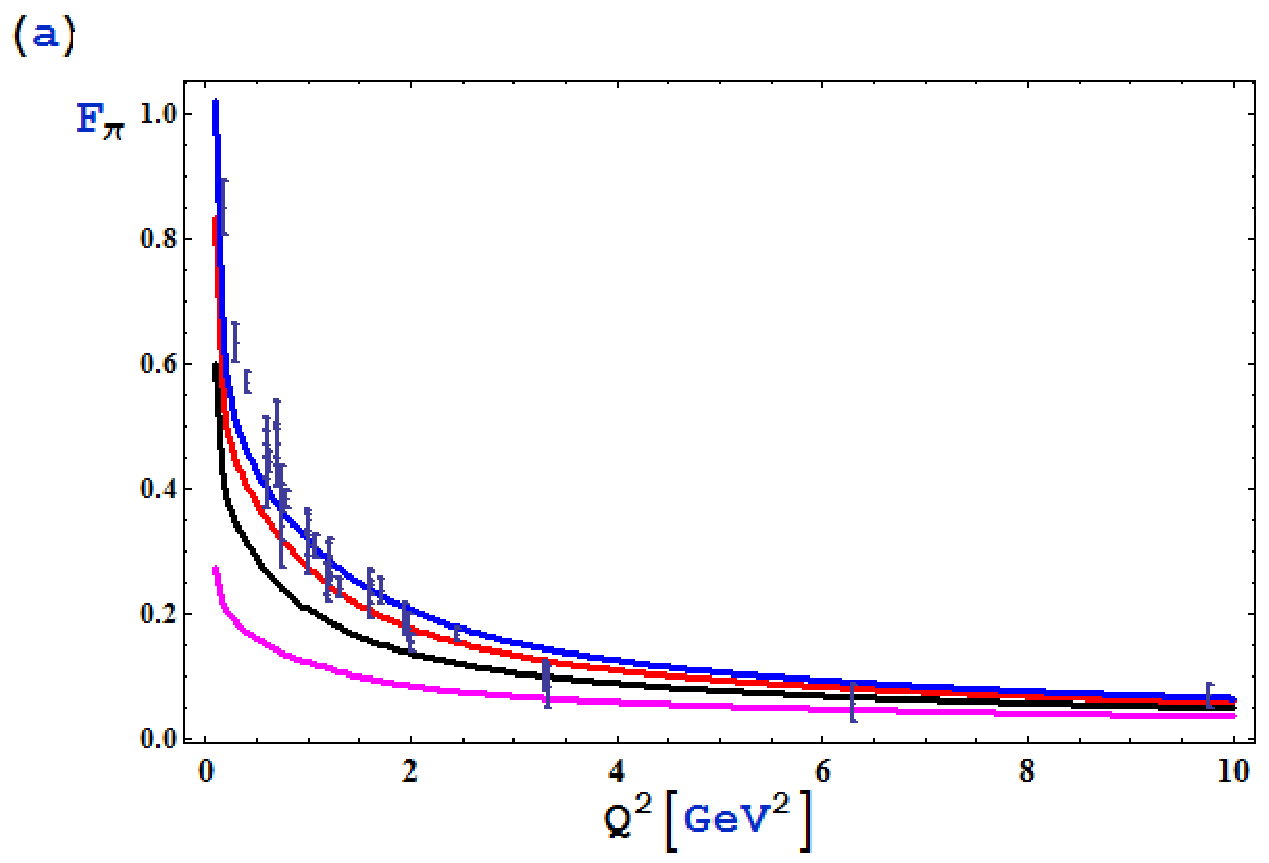}
\includegraphics[width=0.43\textwidth]{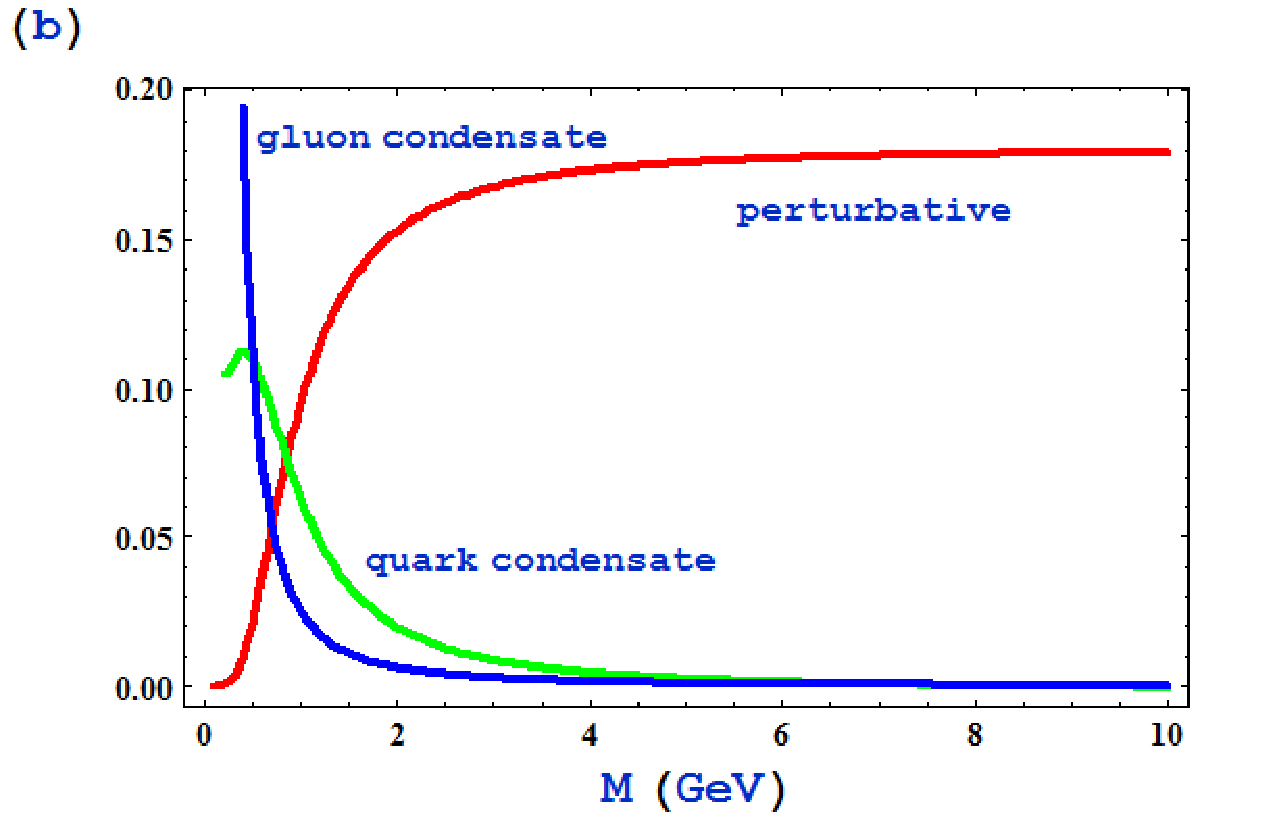}
\caption{(a) $Q^2$ dependence of $F_\pi$ for, from top to bottom,
$\lambda_q=0.8$, 0.75, 0.7 and 0.63 GeV with $c=0.3$. The data
points are referred to \cite{BBHk:78a,Vol:01a,Hor:06a,Tad:06a}. (b)
$M$ dependence of the perturbative and condensate contributions to
$F_\pi(Q^2=1.99\;\mathrm{GeV}^2)$.}\label{expdatafit}
\end{figure}

Our results of the pion form factor $F_\pi(Q^2)$ are displayed in
Fig.~\ref{expdatafit}(a) for three values of $\lambda_q=0.8$, 0.75,
and 0.7 GeV with $c=0.3$, corresponding to the curves from top to
bottom, respectively. Their difference indicates the theoretical
uncertainty of our analysis. It is obvious that all three curves are
well consistent with the experimental data for $Q^2>1$ GeV$^2$, the
region where QSR are applicable. Inputting a smaller value of
$\lambda_q\simeq0.63$ GeV \cite{BPS:09,BI82}
into our formalism directly, a curve lower than
the data is obtained as shown in Fig.~\ref{expdatafit}(a). However,
if increasing the parameter $c$ accordingly up to $c=0.39$
for this different $\lambda_q$, the result will become consistent with the data.
We investigate the perturbative and condensate
contributions to the pion form factor
$F_\pi(Q^2=1.99\;\mathrm{GeV}^2)$ at different Borel mass $M$, as
exhibited in Fig.~\ref{expdatafit}(b). It is observed that the former
increases with $M$, and the latter decreases with $M$ for $M>1$ GeV.
The gluon condensate contribution becomes negligible for $M>1$ GeV,
justifying the sole modification from the nonlocal quark
condensates. Although the magnitudes of different contributions vary
with $M$, their sum is almost constant for $M>1.5$ GeV. The quark
condensates contribute 23\% of the pion form factor
$F_\pi(Q^2=1.99\;\mathrm{GeV}^2)$ at $M=1.5$ GeV, which is slightly
higher than the percentage 17\% in the localization approximation
\cite{IS:82}.

In this letter we have included the nonlocal quark condensates into
QSR via the KL parametrization for a dressed fermion propagator,
which is decomposed into the perturbative and nonperturbative
pieces. The negative spectral density function implies that the
contribution from higher effective quark masses is nonperturbative.
The parametrization of the spectral density functions leads to the
known exponential ansatz for the nonlocal condensate model in our
formalism. We have analyzed the pion form factor as an example, and
the results are in good agreement with the data for $Q^2$ between
1-10 GeV$^2$. The fitted ranges of the virtuality
$\lambda_q$ and of the duality interval $s_0(Q^2)$,
presented in Fig.~\ref{s0choice}, are also consistent with those
reported in the literature. The
nonlocal quark condensates remedy the improper dependence of the
nonperturbative contribution in the localization approximation at
large $Q^2$: the quark condensate effects decrease like $1/Q^2$,
which is different from the exponential decrease obtained in the
literature. Viewing the success of this approach to the pion form
factor, we shall extend it to more complicated processes, including
Compton scattering and two-photon hadron production \cite{HL04}.


We thank A. Khodjamirian for useful discussions during the KITPC
program of Advanced Topics on Flavor Physics in July, 2008. This
work was supported by the National Center for Theoretical Sciences
and National Science Council of R.O.C. under Grant No.
NSC-98-2112-M-001-015-MY3.


\begin{thebibliography}{99}


\bibitem{SVZ:79a} M.A. Shifman, A.I. Vainshtein and V.I. Zakharov, Nucl.
Phys. {\bf B147}, 385, 448, 519 (1979).
\bibitem{Ra:0101} A.V. Radyushkin, arXiv:hep-ph/0101227.
\bibitem{IS:82} B.L. Ioffe and A.V. Smilga, Phys. Lett.
{\bf B114}, 353 (1982).
\bibitem{NR82} V.A. Nesterenko and A.V. Radyushkin, Phys. Lett.
B {\bf 115}, 410 (1982).
\bibitem{BPS:09} A.P. Bakulev, A.V. Pimikov and N.G. Stefanis, Phys.
Rev. D {\bf 79}, 093010 (2009).
\bibitem{CRS93} C. Corian\`{o}, A. Radyushkin, and G. Sterman, Nucl.
Phys. {\bf B405}, 481 (1993);  C. Corian\`{o}, H-n. Li, and C.
Savkli, JHEP {\bf 9807}, 008 (1998).
\bibitem{BM94a} A.P. Bakulev and S.V. Mikhailov, JETP Lett.
{\bf60}, 150 (1994).
\bibitem{MR:86a} S.V. Mikhailov and A.V. Radyushkin, JETP Lett. {\bf43}, 712 (1986).
\bibitem{BM:95a} A.P. Bakulev and S.V. Mikhailov, Z. Phys. C {\bf 68}, 451 (1995).
\bibitem{MR:89a} S.V. Mikhailov and A.V. Radyushkin, Sov. J. Nucl.
Phys. {\bf 49}, 494 (1989).
\bibitem{MR:92a} S.V. Mikhailov and A.V. Radyushkin, Phys. Rev. D {\bf
45}, 1754 (1992).
\bibitem{Ra:9406} A.V. Radyushkin, arXiv:hep-ph/9406237.
\bibitem{BR:91a} A.P. Bakulev and A.V. Radyushkin, Phys. Lett. B
{\bf 271}, 223 (1991).
\bibitem{BM98} A.P. Bakulev and S.V. Mikhailov, Phys. Lett. B {\bf 436},
351 (1998); A.P. Bakulev, S.V. Mikhailov, and N.G. Stefanis, Phys.
Lett. B {\bf 508}, 279 (2001); ibid. B {\bf 590}, 309(E) (2004).
\bibitem{IZbook} G. K$\ddot{\mathrm{a}}$ll$\acute{\mathrm{e}}$n,
Helv. phys. Acta {\bf 25}, 417 (1952);
H. Lehmann, Nuovo Cimento {\bf 11}, 342 (1954).
\bibitem{AS:00a}  R. Alkofer and L.V. Smekal, Phys. Rept. {\bf
353}, 281 (2001).
\bibitem{BM:0203} A.P. Bakulev and S.V. Mikhailov, Z. Phys. C {\bf 68}, 451 (1995);
Mod. Phys. Lett. A {\bf 11}, 1611 (1996); Phys. Rev. {\bf D65},
114511 (2002).
\bibitem{Neu:92a} Matthias Neubert, Phys. Rev. D {\bf
46}, 1076 (1992).
\bibitem{BBBD:92a} E. Bagan, Patrical Ball, V.M. Braun and H.G. Dosch, Phys. Lett. B
{\bf 278}, 457(1992).
\bibitem{Ra:91a} A. V. Radyushkin, Phys. Lett. B
{\bf 271}, 218(1991).
\bibitem{Gr:95} A.G. Grozin, Int. J. Mod. Phys. A {\bf
10}, 3497 (1995).
\bibitem{Ma:84a} S. Mallik, Nucl. Phys. {\bf B234}, 45 (1984).
\bibitem{Co:93a} C. Corian\`{o}, Nucl. Phys. {\bf B410}, 90 (1993).
\bibitem{Io:05} B.L. Ioffe, Prog. Part. Nucl. Phys. {\bf56}, 232 (2006).
\bibitem{BBHk:78a} C.J. Bebek et al., Phys. Rev. D {\bf
17}, 1693 (1978).
\bibitem{Vol:01a} J. Volmer et al., Phys. Rev. Lett. {\bf
86}, 1713 (2001).
\bibitem{Hor:06a}  T. Horn et al., Phys. Rev. Lett. {\bf
97}, 192001 (2006).
\bibitem{Tad:06a}  V. Tadevosyan et al., Phys. Rev. C {\bf
75}, 055205 (2007).
\bibitem{BI82} V.M. Belyaev and B.L. Ioffe, Sov. Phys. JETP {\bf 56},
493 (1982) [Zh. Eksp. Teor. Fiz. {\bf 83}, 876 (1982)].
\bibitem{PW:96} M.V. Polyakov and C. Weiss, Phys. Lett. B
{\bf 387}, 841(1996).
\bibitem{DEM97} A.E. Dorokhov,  S.V. Esaibegyan, and S.V. Mikhailov,
Phys. Rev. D {\bf 56}, 4062 (1997).
\bibitem{HL04} R.C. Hsieh and H-n. Li, Phys. Rev. D {\bf 70},
056002 (2004).





\end{thebibliography}
\end{document}